%% file: main.tex
\documentclass[journal=nalefd,manuscript=letter]{achemso}
\usepackage{color}
\usepackage{lipsum,amsmath}
\usepackage{pifont}   
\usepackage{graphicx} 
\usepackage{dcolumn}  
\usepackage{bm}       
\usepackage{amsfonts} 
\usepackage{amssymb}  
\usepackage{multirow} 
\usepackage{natbib}
\usepackage{multicol}
\usepackage{tikz}
\usepackage{siunitx}
\usepackage{placeins}
\usepackage{braket}
\usepackage{nicefrac}
\usepackage{bm}
\usepackage{physics}
\usepackage{grffile}

    \renewcommand{\v}[1]{\bm{\mathrm{#1}}}
    
    \newcommand{\tx}[1]{\text{#1}}

\input dft

\usepackage[english]{babel}
\usepackage[autostyle, english = american]{csquotes}
\MakeOuterQuote{"}

\title{Ultrafast optical control over spin and momentum in solids}

\author{Q.~Z. Li}
\affiliation{Max-Born-Institut  f\"ur Nichtlineare Optik und Kurzzeitspektroskopie, Max-Born-Strasse 2A, 12489 Berlin, Germany.}
\author{S. Shallcross}
\affiliation{Max-Born-Institut  f\"ur Nichtlineare Optik und Kurzzeitspektroskopie, Max-Born-Strasse 2A, 12489 Berlin, Germany.}
\author{J.~K. Dewhurst}
\affiliation{Max-Planck-Institut f\"ur Mikrostrukturphysik, Weinberg 2, D-06120 Halle, Germany.}
\author{S. Sharma}
\affiliation{Max-Born-Institut  f\"ur Nichtlineare Optik und Kurzzeitspektroskopie, Max-Born-Strasse 2A, 12489 Berlin, Germany.}
\email{sharma@mbi-berlin.de}
\author{P. Elliott}
\affiliation{Max-Born-Institut  f\"ur Nichtlineare Optik und Kurzzeitspektroskopie, Max-Born-Strasse 2A, 12489 Berlin, Germany.}

\date{\today}

\begin{document}

\maketitle

\section{Introduction}

{\bf The coupling of laser light to matter can exert sub-cycle coherent control over material properties, with optically induced currents and magnetism shown to be controllable on ultrafast femtosecond time scales. Here, by employing laser light consisting of both linear and circular pulses, we show that charge of specified spin and crystal momentum can be created with precision throughout the first Brillouin zone. Our hybrid pulses induce in a controlled way both adiabatic intraband motion as well as vertical interband excitation between valence and conduction bands, and require only a gapped spin split valley structure for their implementation. This scenario is commonly found in the 2d semi-conductors, and we demonstrate our approach with monolayer WSe$_2$. We thus establish a route from laser light to local control over excitations in reciprocal space, opening the way to the preparation of momenta specified excited states at ultrafast time scales.}

Sub-cycle control of electrons in matter has already led to the experimental observation of femtosecond control over optically induced current\cite{schiffrin_optical-field-induced_2013,higuchi_light-field-driven_2017,chen_stark_2018}, valley polarization in transition metal dichalcogenides (TMDC)\cite{mak_control_2012,zeng_valley_2012}, and the prediction from first-principles of control over magnetic order faster than the exchange and spin-orbit times\cite{elliott_ultrafast_2016,dewhurst_laser-induced_2018}, subsequently confirmed by several experiments\cite{siegrist_light-wave_2019,steil-heusler,hofherr-feni,clemens-copt,chen_competing_2019}. The crucial limiting factor in fully implementing a coherent lightwave electronics lies in the excited spin and valley state lifetimes, in turn determined by the fundamental scattering processes in materials: the electron-electron scattering, electron-phonon scattering, and spin scattering due to spin-orbit interaction.

The decisive role in such scattering processes is played by crystal momentum, the quantum number associated with periodicity in solids. This determines "hot spots" in the band structure at which the amplitudes associated with electron-phonon interaction or spin relaxation are large. Pre-selecting the crystal momentum of excited states by designed laser pulses thus represents a key step towards designing long-lived valley and spin excitations. However, to date precise lightwave control over the crystal momentum by light has not been demonstrated.

Laser pulses generate two distinct types of excitation in solids: intraband adiabatic transitions in which the crystal momentum $\bf k$ evolves according to the vector potential as $\v k \to \v k + \v A(t)$, and diabatic transitions in which charge is excited between bands. In TMDC semi-conductors with broken inversion symmetry the diabatic transitions induced by circularly polarized light are foundational for valleytronics. However, such excitation is by construction restricted to the high symmetry points (e.g. K and K' points in TMDC), precluding control over crystal momentum. Circumventing this requires a new approach. In this work we demonstrate that hybrid laser pulses combining single cycle terahertz linear light with circularly polarized light allows control over both adiabatic and diabatic motion, facilitating a full control over the crystal momentum. This unprecedented control over electrons in solids by lightwaves opens new possibilities both for the creation of spin and valley excited states, as well as probing the fundamental scattering processes of solids.


\section{Results}

\begin{figure}[t]
\includegraphics[width=.95\textwidth]{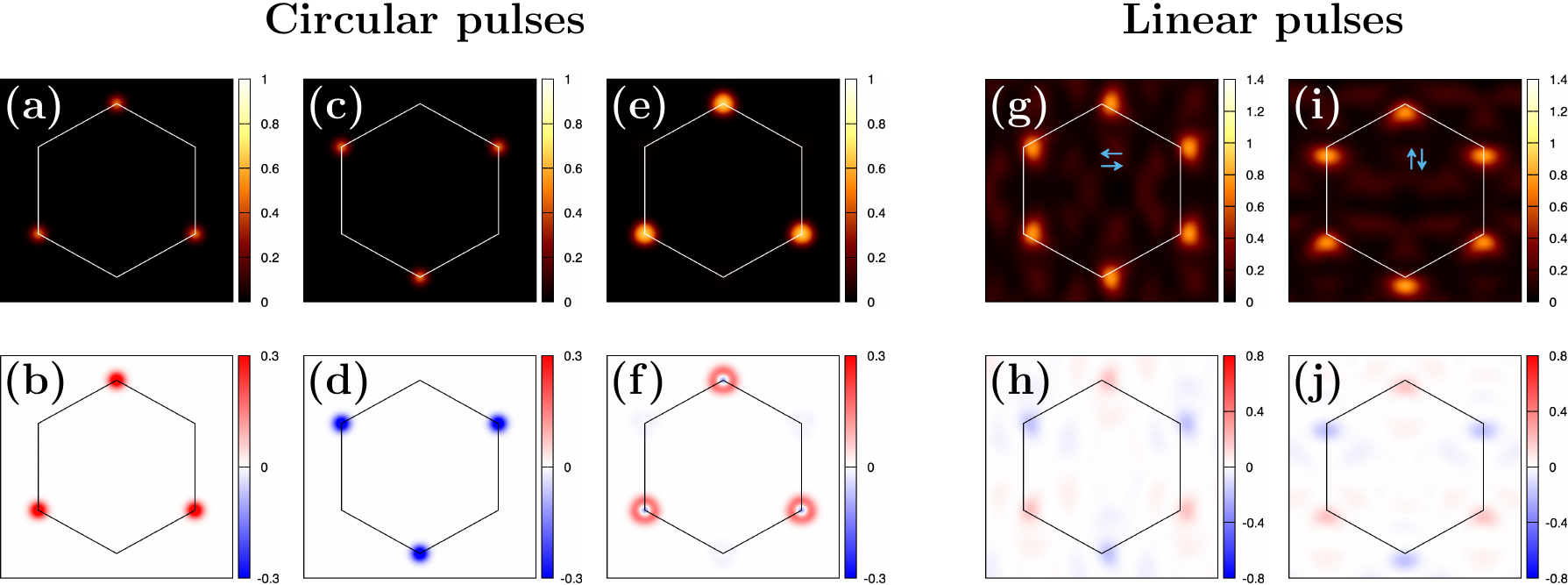}
\caption{\footnotesize{{\it Charge excitation due to circularly and linearly polarized light.} The upper (lower) panels display the charge (spin) excitation in momentum space. A 1.6~eV pulse of circularly polarized light generates pure spin excitation at the verticies of the Billouin zone: (a,b) $\sigma^+$ polarized circular light excites spin up states at the K valley, while (d,e) $\sigma^-$ polarized  light excites spin down states at the K' valley. (e,f) A higher frequency 2.0~eV pulse is in resonance both the spin split valence bands, resulting in a mixed spin but valley specific excitation consisting of (i) spin down excitation at K along with (ii) a surrounding halo of spin up charge due to transitions from the higher energy spin up manifold. In panels (g-j) a single pulse of linear light sufficient to excite interband charge results in asymmetric but a net spin unpolarized excitation, with in each panel the arrows illustrating the intraband motion induced due to: (g,h) $-x$ polarized light, (i,j) $+y$ polarized light.
}}
\label{f:lincirc}
\end{figure}


\begin{figure}[t]
\includegraphics[width=\textwidth]{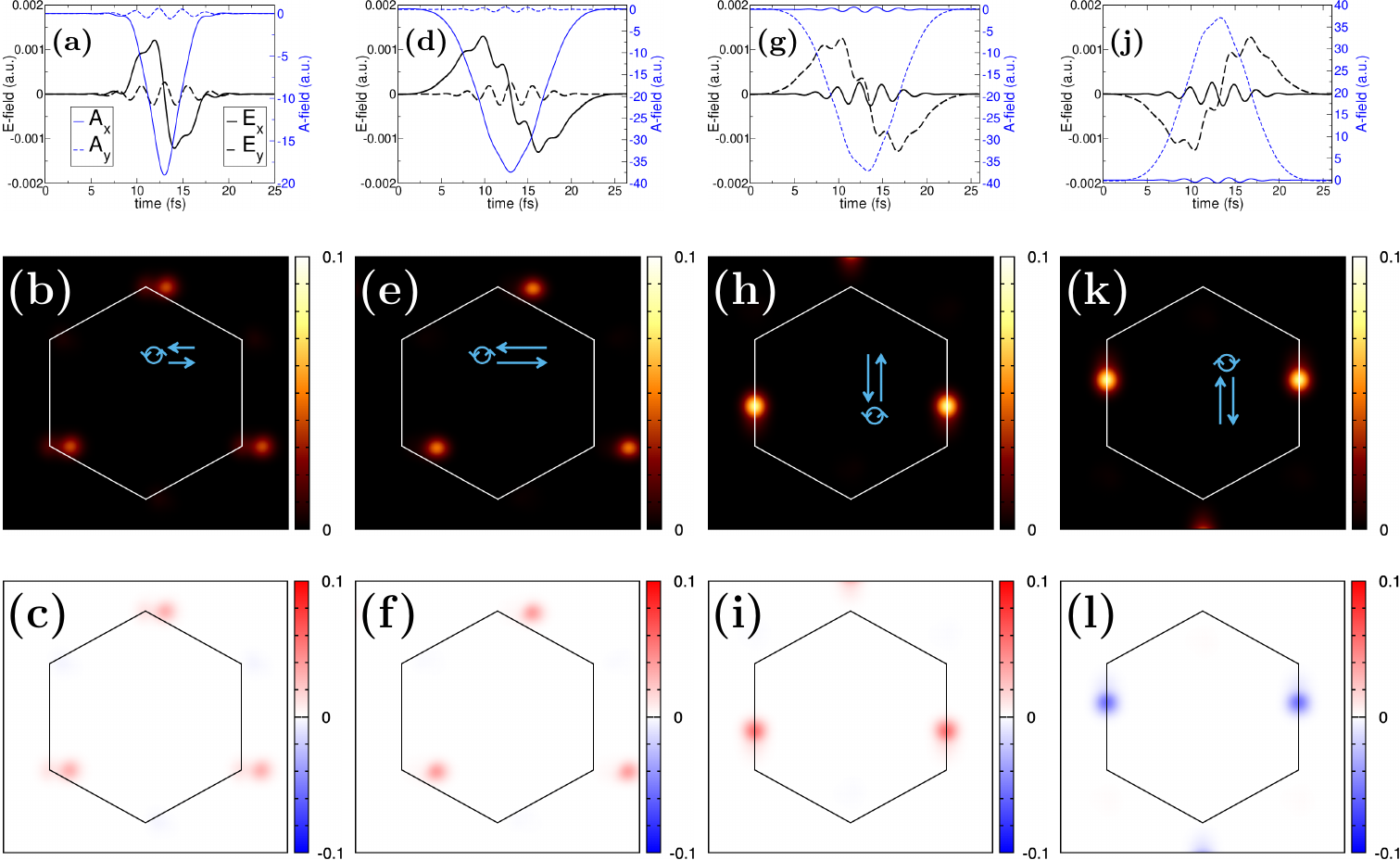}
\caption{\footnotesize{Laser control over both crystal momentum and spin polarization in WSe$_2$. A combination of a weak THz linearly polarized pulse with, at half cycle, a circularly polarized optical pulse generates charge excitation at pre-selected crystal momenta. With $-x$ polarized linear light and $\sigma^-$ polarized circular light this hybrid pulse generates spin up excitation displaced from the K valley in the $k_x$-direction by an amount depending on the amplitude of the linear pulse, see panels (a-c) and (d-f). A $-y,\sigma^-$ hybrid pulse generates spin up polarization in the vicinity of the M point, panels (g-i), while a $+y,\sigma^+$ hybrid pulse results in spin down polarization in the vicinity of the M point, panels (j-l). In each panel the blue arrows denote the type of hybrid pulse with the pulse vector and scalar potentials displayed in the topmost panels.}}.
\label{f:IRcirc}
\end{figure}

Due to broken inversion symmetry and strong spin-orbit coupling WSe$_2$ presents an ideal case for the implementation of this idea. At the K/K' valleys this material has significant spin spitting (a maximum of $450$~meV\cite{xiao_coupled_2012} at these K/K' points) in addition to a gap of 1.6~eV. This allows for $\sigma^+$/$\sigma^-$ circularly polarized light in the optical range to excite spin up/down polarized charge at the K/K' valleys, so called valley-spin locking. This is demonstrated, by the means of a state-of-the-art first principles calculation performed using time-dependent-density functional theory (TDDFT), in Figs.~1a-d (for details of the computational technique see Methods section). A pump laser pulse with frequency equal to the gap was used (for other pulse parameters see Ref.~\cite{pulse}). Increasing the pulse frequency brings both bands into resonance but at different crystal momenta, generating a mixed spin excited charge density consisting of the lower band spin at the valley centre surrounded by a halo of the higher band spin, see Fig.~1e,f. For sufficiently strong linear light fields, the intraband motion induced by the pulse is accompanied by interband excitation via Landau-Zener (LZ) tunneling. This is shown in Fig.~1g-j for two different polarizations of linear light. LZ tunneling is effective at all valleys, and leads to a somewhat asymmetric valley excitation, that depends on the trajectory in reciprocal space induced by the linear light. The excited states generated by linear and circular pulses are thus restricted to the high symmetry points of the Brillouin zone.

On the other hand, the presence of the gap allows for sufficiently weak linear light to adiabatically evolve Bloch states according to the acceleration theorem $\v k(t) = \v k(t=0) + \v A(t)$ without exciting interband Landau-Zener transitions. Thus, by employing a single cycle pulse we can propagate a state of arbitrary crystal momentum $\v k(t=0)$ on the valence band manifold to one of the high symmetry K/K' points and back to $\v k(t=0)$. If, however, at exactly the centre of this pulse (i.e., at half-cycle where the {\bf A} field is maximal) we apply circularly polarised light, this state will be excited to the conduction band, before returning under the action of the linear light to $\v k(t=0)$. Thus, charge will have effectively been excited vertically $\v k(t=0)$ via a path through the "hot spots" at K/K'. Our hybrid pulse therefore consists of a infra-red linear pulse in combination with optical circularly polarized light.

The IR component of this hybrid pulse can control the displacement from the valley centre of the excited charge. To demonstrate this we first consider a IR Terahertz pulse centered on frequency $0.4$eV ($97$THz) with fluence $0.2$ mJ/cm$^2$ (see Fig.~\ref{f:IRcirc}a), generating the localized charge excitation shown in Figs.~\ref{f:IRcirc}b,c. If we now consider a second pulse with double the vector amplitude with, in order to prevent LZ transitions, the frequency is reduced to $0.2$eV ($48$THz), causing the fluence to increase to $0.4$ mJ/cm$^2$, the excited charge is approximately double the distance from the K' point. Note that the peak intensity is almost the same for both pulses and both are linear polarized in $-x$ direction. Thus by tailoring the frequency and polarization of the IR pulse, we can excite spin at a pre-determined point in the 1BZ in principle arbitrarily far from the high symmetry K points.

The circular component of the pulse, on the other hand, offers control over the spin-polarization of the excited charge. To demonstrate this we look at the region around an M point and excite both up and down spin. In Fig.~\ref{f:IRcirc}g-i, we use the same IR pulse as Fig.~\ref{f:IRcirc}a-f but instead oriented with $-y$ polarization. This takes the M point electrons down to the K point where a $\sigma^+$ pulse excites spin-up polarization, seen in Fig \ref{f:IRcirc}e,f. Conversely in Figs.~\ref{f:IRcirc}j-l, we now change the linear polarization of the IR pulse to the $+y$ direction and deploy a $\sigma^-$ pulse at half-cycle. This brings the M point up to the K' point and excites spin-down electrons. Thus, while both choices excite the M-point region, by tailoring the pulses we can decide whether spin-up or spin-down electrons are excited.\\


\begin{figure}[t]
\includegraphics[width=0.80\textwidth]{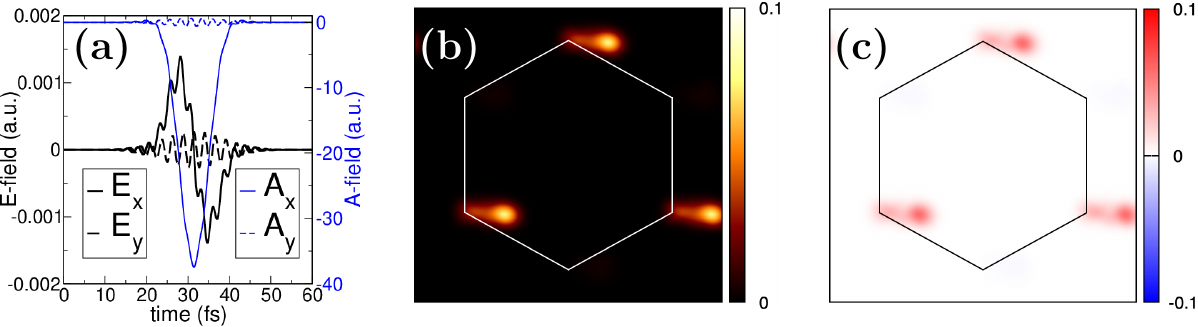}
\caption{\footnotesize{Impact of the width of the circular pulse on crystal momentum and spin polarization controlled charge excitation in monolayer WSe$_2$. By increasing the duration of the circular pulse charge excitation at the high symmetry K point occurs for a increased proportion of the intra-band trajectory, resulting in a comet like charge excitation. This should be compared with Fig.~\ref{f:IRcirc} in which a tighter circlar pulse results in a more focused charge excitation}}.
\label{f:IRcircSLOW}
\end{figure}

Crucial to the controlled excitation of momentum in ${\bf k}$-space described thus far is that the circular pulse occurs at half cycle of the weak IR linear pulse. Evidently, to create such a hybrid pulse the width of the circular pulse must be narrow as compared to that of the linear pulse. The circular pulses considered thus far have a FWHM of 7.5~fs, an experimentally challenging laser field although one within reach of present day pump laser capability. We thus now consider how increasing the width of the circular pulse to a standard 20fs duration impacts the charge excitation (this laser pulse is shown in Fig.~\ref{f:IRcircSLOW}a). We would expect that a longer duration of inter-band excitation in relation to the intraband motion to result in the smearing out of the focused charge excitation into a continuous scar, and indeed as seen in Fig.~\ref{f:IRcircSLOW}b-c we see that the intense localized charge excitations seen in Fig.~\ref{f:IRcirc} takes on a "comet" like appearance.\\

\section{Discussion}

Future ultrafast nano-technology relies upon precise control over the electronic degrees of freedom of materials through designed laser pulses. In the present work we demonstrate such a control; carefully sculpted light pulses, combining weak linearly polarized pulses with  circularly polarized pulses, offers full control over excitations with specified spin and crystal momentum. The materials which offer such a control require a gapped spin-split valley-type band structure which is a common feature amongst 2d semiconductors. As the magnitude of this gap is generally controllable via tuning by a layer perpendicular electric field (the giant stark effect\cite{kim_ultrafast_2014}) the full control over spin and crystal momentum we describe here should therefore be possible in many 2d materials. Furthermore, we show that this physics occurs at ultrafast time scales, fully controlled by the pulse, and before electron-electron and electron-phonon scattering processes. Such full control, in addition to representing a hitherto unexplored richness of ultrafast phenomena in 2d materials, will be useful in probing momentum and spin dependent scattering processes that will occur post charge excitation, such as electron-phonon coupling, and will further also be useful in tailoring the momentum character of injected charge in an interface geometry.

\section{Methods}

The Runge-Gross theorem \cite{RG84} establishes the time-dependent external potential as a unique functional of the time dependent density, given the initial state. Based on this theorem, a system of non-interacting particles can be chosen such that the density of this non-interacting system is equal to that of the interacting system for all times\cite{EFB09,C11,SDG14}, with the wave function of this non-interacting system represented by a Slater determinant of single-particle orbitals. In a fully non-collinear spin-dependent version of these theorems\cite{KDES15} time-dependent Kohn-Sham (KS) orbitals are Pauli spinors governed by the Schr\"odinger equation:

\begin{eqnarray}
i\frac{\partial \psi_j({\bf r},t)}{\partial t}&=&
\Bigg[
\frac{1}{2}\left(-i{\nabla} +\frac{1}{c}{\bf A}_{\rm ext}(t)\right)^2 +v_{s}({\bf r},t) + \frac{1}{2c} {\sigma}\cdot{\bf B}_{s}({\bf r},t) + \nonumber \\
&&\frac{1}{4c^2} {\sigma}\cdot ({\nabla}v_{s}({\bf r},t) \times -i{\nabla})\Bigg]
\psi_j({\bf r},t)
\label{KS}
\end{eqnarray}
where ${\bf A}_{\rm ext}(t)$ is a vector potential representing the applied laser field, and ${\sigma}$ are the 
Pauli matrices. The KS effective potential $v_{s}({\bf r},t) = v_{\rm ext}({\bf r},t)+v_{\rm H}({\bf r},t)+v_{\rm xc}({\bf r},t)$ is decomposed into the external potential $v_{\rm ext}$, the classical electrostatic Hartree potential $v_{\rm H}$ and the exchange-correlation (XC) potential $v_{\rm xc}$. Similarly the KS magnetic field is written as ${\bf B}_{s}({\bf r},t)={\bf B}_{\rm ext}(t)+{\bf B}_{\rm xc}({\bf r},t)$ where ${\bf B}_{\rm ext}(t)$ is the magnetic field of the applied laser pulse plus possibly an additional magnetic field and ${\bf B}_{\rm xc}({\bf r},t)$ is the exchange-correlation (XC) magnetic field. The final term of Eq.~\eqref{KS} is the spin-orbit coupling term. It is assumed that the wavelength of the applied laser is much greater than the size of a unit cell and the dipole approximation can be used i.e. the spatial dependence of the vector potential is disregarded. All the implementations are performed using the state-of-the art full potential linearized augmented plane wave (LAPW) method. Within this method the core electrons are treated fully relativistically by solving the radial Dirac equation while higher lying electrons are treated using the scalar relativistic Hamiltonian in the presence of the spin-orbit coupling. To obtain the 2-component Pauli spinor states, the Hamiltonian containing only the scalar potential is diagonalized in the LAPW basis: this is the first variational step. The scalar states thus obtained are then used as a basis to set up a second-variational Hamiltonian with spinor degrees of freedom \cite{singh}. This is more efficient than simply using spinor LAPW functions, however care must be taken to ensure that a sufficient number of first-variational eigenstates for convergence of the second-variational problem are used.
A fully non-collinear version of TDDFT as implemented within the Elk code\cite{elk} is used for all calculations.

To calculate the crystal momentum, {\bf k}, resolved excitation we use the expression

\begin{equation}
\label{nex}
N_{\tx{ex}}(\v k) = \sum_a^{occ}\sum_b^{unocc} \left|\braket{\psi_{a\v k}(t)}{\psi_{b\v k}(t=0)}\right|^2
\end{equation}
Here the time-dependent KS orbitals at a given time $t$ are projected on to the ground-state orbitals to calculate the change in occupation of the KS orbitals. Formally, within TDDFT, the transient occupation of the excited-states does not necessarily follow that of the KS system. For weakly excited systems, however, the difference is expected to be small. It should be noted that for high fluence pulses where the band renormalization effects are large, not applicable to this work, such an approximation would fail.

\section{Acknowledgements}
QZL would like to thank DFG for funding through TRR227 (project A04). SS would like to thank DFG for funding through SH498/4-1 and PE acknowledges funding from DFG Eigene Stelle project 2059421. The authors acknowledge the North-German Supercomputing Alliance (HLRN) for providing HPC resources that have contributed to the research results reported in this paper.

\bibliography{Paper11}

\end{document}

%% file: dft.tex
\def\bea{\begin{eqnarray}}
\def\eea{\end{eqnarray}}
\def\ben{\begin{equation}}
\def\een{\end{equation}}
\def\benu{\begin{enumerate}}
\def\enu{\end{enumerate}}
\def\bei{\begin{itemize}}
\def\eei{\end{itemize}}

\def\sss{\scriptscriptstyle\rm}

\def\1var{(\bx_1...\bx\N)}

\def\br{{\bf r}}

\def\bx{{\br t}}

\def\N{_{\sss N}}

\def\sph_int{ {\int d^3 r}}


%% file: main.bbl
\begin{mcitethebibliography}{23}
\bibitem[pul()]{pulse}
The circular pulses have full width half maximum $7.26$ fs and fluence $0.138$
  mJ/cm$^2$ for both the $\omega=1.6$ eV and $\omega=2.0$ eV pulse. The linear
  polarization cases consist of single cycle pulses centered on frequency $2.5$
  eV and fluence 0.84mJ/cm$^2$. The vector potential has a simple gaussian
  shape, $A(t) = 18.41 e^{-0.002 t^2}$, in atomic units.\relax
\mciteBstWouldAddEndPunctfalse
\mciteSetBstMidEndSepPunct{\mcitedefaultmidpunct}
{}{\mcitedefaultseppunct}\relax
\EndOfBibitem
\end{mcitethebibliography}
